\documentclass[12pt, draftclsnofoot, onecolumn]{IEEEtran}
\IEEEoverridecommandlockouts
\usepackage{cite}
\usepackage{amsmath,amssymb,amsfonts}
\usepackage{algorithmic}
\usepackage{graphicx}
\usepackage{textcomp}
\usepackage{xcolor}
\usepackage{subfigure}
\usepackage[utf8]{inputenc}

\def\BibTeX{{\rm B\kern-.05em{\sc i\kern-.025em b}\kern-.08em
    T\kern-.1667em\lower.7ex\hbox{E}\kern-.125emX}}
\begin{document}

\title{Critical Checkpoints for Evaluating Defence Models Against Adversarial Attack and Robustness}

\author{\IEEEauthorblockN{Kanak Tekwani}\\
\IEEEauthorblockA{\textit{Institute of Technology, Nirma University} \\
Ahmedabad, India \\
19bec134@nirmauni.ac.in}\\
\and
\IEEEauthorblockN{Manojkumar Parmar}\\
\IEEEauthorblockA{\textit{Swiss School of Business and Management} \\
Geneva, Switzerland \\
manojkumar.parmar@ssbm.com\\\textit{Bosch Global Software Technologies Pvt. Ltd.}\\Bengaluru, India\\manojkumar.parmar@bosch.com}
}
\maketitle

\begin{abstract}
From past couple of years there is a cycle of researchers proposing a defence model for adversaries in machine learning which is arguably defensible to most of the existing attacks in restricted condition (they evaluate on some bounded inputs or datasets). And then shortly another set of researcher finding the vulnerabilities in that defence model and breaking it by proposing a stronger attack model. Some common flaws are been noticed in the past defence models that were broken in very short time. Defence models being broken so easily is a point of concern as decision of many crucial activities are taken with the help of machine learning models. So there is an utter need of some defence checkpoints that any researcher should keep in mind while evaluating the soundness of technique and declaring it to be decent defence technique. In this paper, we have suggested few checkpoints that should be taken into consideration while building and evaluating the soundness of defence models. All these points are recommended after observing why some past defence models failed and how some model remained adamant and proved their soundness against some of the very strong attacks.    
\end{abstract}


\section{Introduction}
While creating any model of ML or DL it is very crucial to understand that how you can subvert them. Our goal should not only be to train the model but instead train on such robust features so that it cannot be tricked easily.
Giving a rudimentary example if one wants a cat image to be classified as dog, one can easily fool a neural network by perturbing the image that is even imperceptible to human eye. Yes, identifying a cat image as dog would not be a matter of concern to us, but if while identity verification at security check, such small perturbation to image fooling the neural network would not be appreciated.

Similar goes for audio as well, it has been experimented that a well-crafted sound can be generated by such adversaries which is incomprehensible to humans. But if played that same sound to any voice assistant like Apple Siri or Google Assistant, its machine learning algorithm can identify that is saying “Please send all the recent photos of mine to xyz contact” or any similar command. The voice assistant would implement that command and your private information could be in utter danger[1].

Spammers modify their spam messages a little bit get through.

It is easy to interpret the working of linear regression type models, but in past couple decades the rise of complex non-linear algorithm in machine learning have made it difficult to explain that which features the model is taking into consideration while training. One has to go with example based explanation. One has to change the input by small amount and check how the output prediction changes. ML models are significantly less interpretable than statistical models. Interpretability and explainability is of utter importance when building defence model.

So in this paper are some points that any researcher should check when building his defence model. These checkpoint are not sufficient or necessary points that any defence model must check, but in retrospect looking at the past papers, it is evident that if the defence model follows these recommendations, the probability of attackers finding any flaw in defence model significantly reduces. Each time the defence was broken, we got an idea how they broke and what further precautions are needed to be taken by the researchers building novel defence models.
Making a defence model is not as simple as one might think. A defence model has to work against all attack, but to prove it wrong only attack is sufficient. 

 \begin{figure}
     
     \includegraphics[width=85mm,scale=0.7]{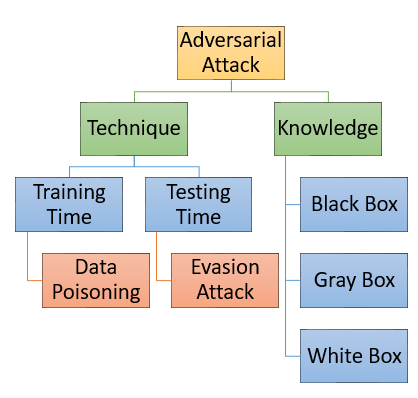}
     \centering
     \caption{Types of attack}
     \label{fig:my_label}
 \end{figure}

\section{What is Adversarial Attack?}
Before understanding how defence should be made robust, we need to understand which different types of attack exist and have taken place. 

An adversarial attack is to subtly perturbing the image in such a way that it is imperceptible to human eyes but the feature space is changed significantly and it deteriorates the performance of classifier.

\subsection{Types of Attack}
\textbf{Data Poisoning Attack:} In this type of attack the malicious person is successful in injecting perturbed image in training set of images. And is capable of shifting the decision boundary. These type of attack have proven to be stronger and have even if attacker is able to poison 3\% of training set it could plummet the accuracy by 11\% [2].
Microsoft’s chatbot was also affected by this. Chatbot was learning from people interacting with it and many people feed it content that was inappropriate. As, a result Micreosft had to shut it down within 24 hours [3].

\textbf{Evasion Attack:} This attack is similar to data poisoning. But it pushes the poisoned data on the other side of the decision boundary so that the classifier misclassifies the data.

\textbf{Whitebox Attack:} Here the malicious person has complete information about the victim’s model architecture and its parameter.

\textbf{Graybox Attack:} Here the malicious person has partial knowledge about the architecture and parameters of the model.

\textbf{Blackbox Attack:} Here the malicious person only knows the output of the victim’s model but has no access to it’s architecture. Here we perturb the image in small amount and we can notice the changes in loss. By observing these two quantities we can have a better idea of numerical gradient.

\section{MATERIALS AND METHOD}

\subsection{What Does Actual Attack Look Like?}\label{AA}
If suppose we are classifying if an image is man or helicopter. Assume our model is fairly classifying with high accuracy. Now we take 1 base image from that dataset and perturb it by adding calibrated amount of noise to it. The noise added to image will be indiscernible to human eyes but is capable enough to fool the neural network. 

Suppose there is a decision boundary, now the poisoned image to human eye looks normal but in feature space it will represent the other class. Here in this attack we know the feature extractor that is been used. But in case if we are devoid of information that which feature extractor is used we can guess it. And we hope that it makes a feature collision with the actual extractor. So here is a fair chance that we miss it. So as a solution to this we can instead of training one poison more number of images(large number of feature space).
We take an image of cat, we take the derivative image with neural network with respect to some loss function. And it will generate a noise, which when added to the image and given to neural network, it can easily be fooled.

The big question is about the transferability of this perturbation models. If one can fool one model using particular perturbation, surprisingly the same perturbation is able to attack another model as well which has different parameters and architecture.

 \begin{figure}
     \centering
     \includegraphics[width=95mm,scale=0.75]{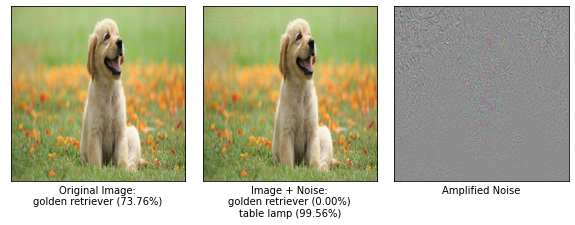}
     \caption{Attack on Inception V3 model}
     \label{fig:my_label}
 \end{figure}
 
 Here we have used inception V3 model to attack(Fig. 2). We have modified the model so it takes input image in 299x299 size. The model classifies this as ‘Golden Retriever’ with 73\% accuracy, but after we added the carefully crafted noise to it, it is predicting ‘table lamp’ which we wanted it to be as this was a targeted attack. Here we kept the noise limit to be 5 so that it is indiscernible to humans. And stopped the iteration if the score reaches 99 percent. First starting with 0 noise and optimizing noise with respect to target class. 
Similar way, we can target the same dog image as an ‘African Elephant’ or anything else.(Fig. 3,4)  

 \begin{figure}
     \centering
     \includegraphics[width=95mm,scale=0.75]{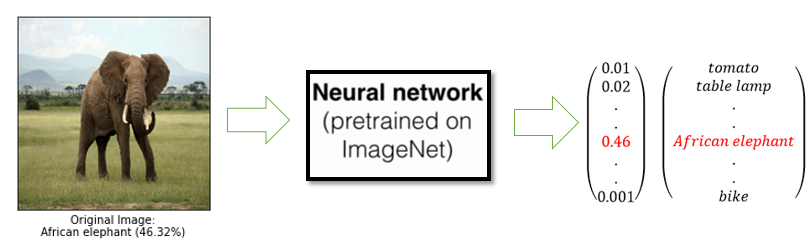}
     \caption{Before Introduction of Calibrated Noise}
     \label{fig:my_label}
 \end{figure}
 
  \begin{figure}
     \centering
     \includegraphics[width=100mm,scale=0.9]{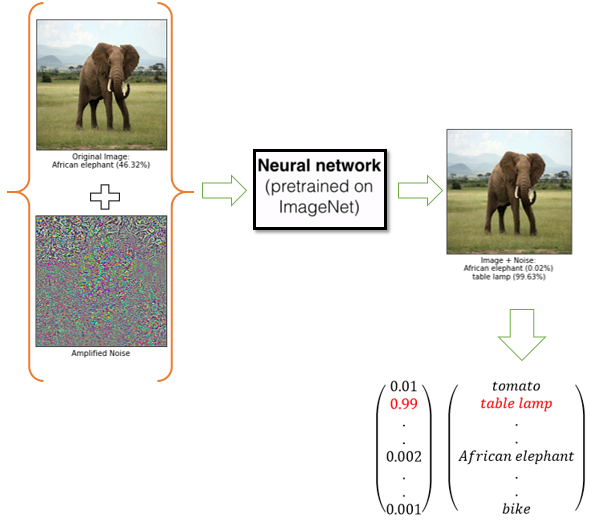}
     \caption{After Introduction of Calibrated Noise}
     \label{fig:my_label}
 \end{figure}

\subsection{Robust and Non-Robust Features}
If we look from neural network perspective the image of dog is just collection of pixels, we have just trained it to increase its accuracy and identify some pixel patterns in image.

Here interpretability of machine learning comes into picture. Robust features with adversaries were correlated with labels, but non-robust features somewhat correlated with labels and can be easily manipulated. Also non-robust features are beyond comprehension to eyes of human[4].

So basically if the convolutional neural network pays attention to more generalized features and learns from these features adversary can easily be injected into it. 

We are getting high robust accuracy i.e. model is performing well even on adversarial test set. So the difference between general and this model is normal classifier are only good getting high accuracy on standard dataset (images without perturbation) but are easily fooled by adversarial images, but robust classifier also works well on perturbed images. As they are trained on robust features.

So after misclassifying the data intentionally, now the robust features have still the original labels, so robust features are misleading and non-robust features are the one that are actually true. But as the model will actually rely on the non-robust feature as it is correlated[5].

So as a solution we can take the whole dataset and restrict the training to just robust features so even if there are perturbations it will not affect and will give robust accuracy. As indirectly non-robust features in data are responsible for adversaries. Non-robust features arguably deteriorates the human mimicking capability of machine learning. 

So the answer to the question that why a particular type of perturbation is valid to attack other different models also, is the neural networks takes into account all the features including robust and non-robust, and the probability is high that these all models are learning on similar non-robust features. But if we train our model on robust features itself, it will help us to overcome the vulnerabilities of our model.

\subsection{Gradient Attack and Defence}
We take the gradient of some network that we have access to and generate a small epsilon perturbation and giving that epsilon perturbation to the dataset and we will be able to classify.

\subsubsection{Gradient Masking}

As a defence of gradient attack we can mask our gradients, so attacker would find it difficult to find the gradients. We can try to make model that always has 0 gradient.

Suppose our model has many class and upon inserting one image it gives top 5 classes that have the highest probability, so the attacker would have these four other classes where he can train the adversary and make it misclassify. So if we present only one class as our output, attacker would find it difficult to perturb the image towards any specified target.

But guess what this zero gradient method also fails to defend our model, as if the output was elephant. Our model will still give output as elephant even after the perturbations were applied. Here there are fairly high chances the attacker might be able to find that these particular points were perturbed by him and the model’s output still didn’t change. So he is aware about the shortcomings of model and now capable of training his own model with gradients, and generate adversarial examples. And there is high probability that these new images will misclassify. So investing in zero gradient method would not be a decent solution to this problem. 

  \begin{figure}
     \centering
     \includegraphics[width=100mm,scale=0.9]{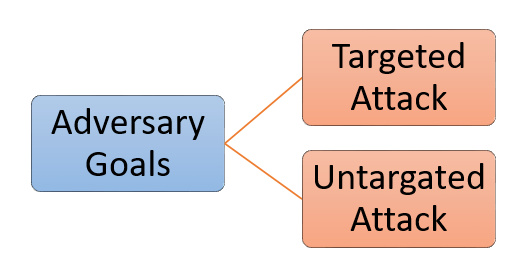}
     \caption{Adversary Goals}
     \label{fig:my_label}
 \end{figure}

\textbf{Targeted Attack}: Here the attacker is interested in misclassifying a source class to a particular target class.

\textbf{Untargeted Attack}: Here attacker does not have any specific target class. They just misclassify the data to some class.
\vspace{5mm}

\subsubsection{Rudimentary Defences}

Normal or benign users are not willing to run many queries. Attackers who prefer test time attack arguably need to run an enormous amount number of queries. As a defence we can limit the access that one can have to model. Only allow user to run a threshold amount of queries so that it would be difficult for attacker to discover weak spots of our model.\\

\subsubsection{Reliability on Assumptions}
New models of machine learnings work great with I.I.D datas. I.I.D stands for independently and identically distributed. Identical here indicates the distribution of the data that is generated does not change over time rather stays fixed. Independent here indicates the sample that we take is independent of the sample that we are going to take next. When we tend to drop this assumptions the model’s performance plummets.

Dataset shift occurs when training and test distributions are different. This datashift occurs due various activation function in subsequent layers of neural network, numerous ways of selecting testing and training sets, sparsity of data. Suppose when we have train and test dataset where the relative proportion of data points differ a lot in space. So the model has to do some sacrifices and this leads to covariate shift. Also when trained most of the dataset on augmented images, the classifier becomes more sensitive to even single pixel perturbations. 

Most researchers focus on part when the malicious person tries to violate the assumption of identical criteria. Not much importance is given to independence criteria. Researchers should try a model that changes each time it runs similar to dynamic model for security purpose.

\subsection{Threat Models}

Threat model characterizes when the defence system is intended to be secure. We identify potential short-comings and come up with curative measures so that these short-comings are not exploited. And what actions we take to prevent any loses that might occur due to these vulnerabilities. Here comes goals of adversaries.

Another approach can be from architectural perspective. Suppose for some metric M and natural image x, there is an valid adversarial image x' M(\, x, x') \, $\leqslant$ $\varepsilon$ if x' is misclassified and $\varepsilon$ is very small(\,if we perturb the image more it is obvious we can misclassify the image)\,. This metric is made by considering the fact that such small change of $\varepsilon$ will not change the class of input and if some perturbation here makes the model to misclassify than it would be a matter of concern. Now the question arises what M do we assume here…

Generally for images Lp norm is used. For example L$\infty$ norm for $\varepsilon$ constraint adversary cannot perturb any specific pixel by more than $\varepsilon$ amount. However for different purposes different choice of d metric and value of $\varepsilon$ is considered. No matter which metric one uses it is nearly implausible to exactly calculate the evaluation, these results are always approximate. If one’s defence is not working well against these norms, the probability that it will not be robust in real application highly increases.

If any defence paper claims to be defensible against L$\infty$ norm for less than 0.3, it should not assume that the defence will also work fine for L2 norm for any other value.

\subsection{What Possibly Can Break The Defence?}
After proving that the previously existing attacks fail to break one’s novel defence model, the paper should provide the vulnerabilities of their defence model too. After confirming that proposed defence model is robust against existing attack models even after conscientiously tuned hyperparameters, it should dive deeper and also check for other some novel attack that attacker might possibly think of generating to break this model. 
According to zero knowledge proof given in [6]\\

It should take care of these three integrity issues(fig. 6):

  \begin{figure}
     \centering
     \includegraphics[width=100mm,scale=0.9]{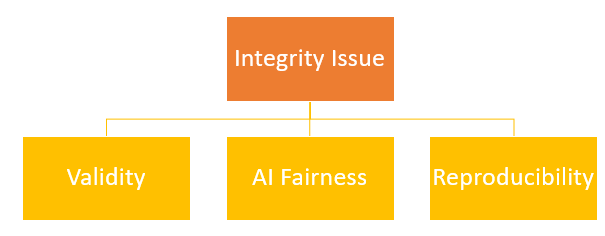}
     \caption{Integrity Issues}
     \label{fig:my_label}
 \end{figure}
 
\begin{itemize}

\item \textbf{Validity}:In many papers authors claim to use some technique, later comes out that it was not actually been used but the result was the consequence of different method.

\item \textbf{AI Fairness}: Concerns are regarding how one can examine if model is biased to some of the features. Minority classes often have lower accuracy. Which predictions should we correct[7]? It is question of high interest if we can validate the model accuracy without revealing the architecture of model, which is answered by zero knowledge proofs. Zero knowledge proof (ZKP) allows you to evaluate the model without revealing the whole architecture of the defence model. It consists of ‘verifier’ and ‘prover’. Prover makes an attempt to prove to the verifier without revealing to verifier about that specific thing. If the prover fails to persuade verifier and verifier feels the information given to it is false the ZKP criteria will not be satisfied. Verifier in this whole process will only learn about the output. 

\item \textbf{Reproducibility}: There are many researches in which it is claimed that their specific model achieves ‘x’ amount of accuracy (evaluated on particular dataset), but the same accuracy is never obtained again by other folks.

\end{itemize}

\subsection{Adaptive Attacks and Defence}

Adaptive attacks should be performed on defence models. It should be assumed attackers have full access to defence model and end to end evaluation should be carried out. 
Graph should be plotted for success rate of attack vs. perturbation budget[8].

Also plot for effectiveness of attack vs. number of iteration graph, as for all attack models it is not necessary that increasing the number of iteration will lead to better attack. Iterative attacks are mostly better than single step attacks, if the iterative attacks is not surpassing single step attack, there is high possibility that iterative attack is wrong. Also graph between accuracy of defence model vs how much perturbation one has injected into it. This graph should be decreasing after some higher value of $\varepsilon$ [9].

  \begin{figure}
     \centering
     \includegraphics[width=100mm,scale=0.9]{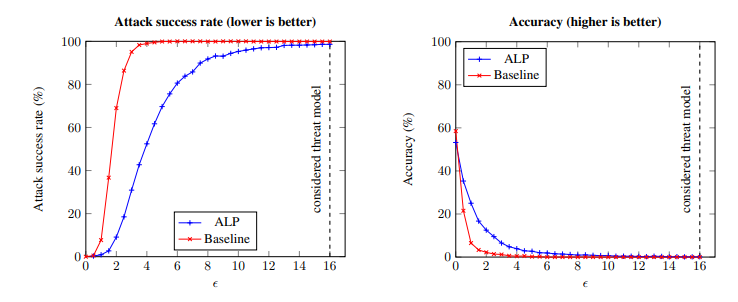}
     \caption{Attack success rate vs. $\varepsilon$ and accuracy vs $\varepsilon$ }
     \label{fig:my_label}
 \end{figure}

\subsection{Defence against Transferability of model}

Suppose attacker has access to training data on which the model is been trained. And he queries the model and gets the prediction on it. That would lead to some classification. And attacker now queries the model by the data outside the training data and he gets another classification.

Now only thing left to do is find difference between the above two obtained classification by running a ML binary classifier model which further will become attack model. 

In the above scenario we considered that attacker has access to training dataset. Now if we consider that the attacker does not have access to training data. He can get the predicted output of the target model and synthesize training data and train it on a model similar to that of attack model. We call such models as shadow model. Attackers train their attack models based on prediction of their shadow model. 

So attacks have versatile nature and can work on different models as well, so a defence model should check for transferability analysis to claim that defence model is strong enough to protect against such versatile nature of attacks.

\subsection{Attack and evaluate}

One line of code that is common is:\\
acc,loss = model.evaluate(x\textunderscore test, y\textunderscore test)

What a strong defence model should evaluate on is:\\
acc,loss = model.evaluate(A(x\textunderscore test), y\textunderscore test)
basically should run an attack on test set. Obtaining the function A would not be as straightforward as it seems. But it makes sure it is evaluated on worst case possible.

\begin{figure}
    \centering
    \subfigure[Fig. 8a This is what the loss function of a standard neural network looks like.]{
    \includegraphics[width=0.5\linewidth,height=6cm]{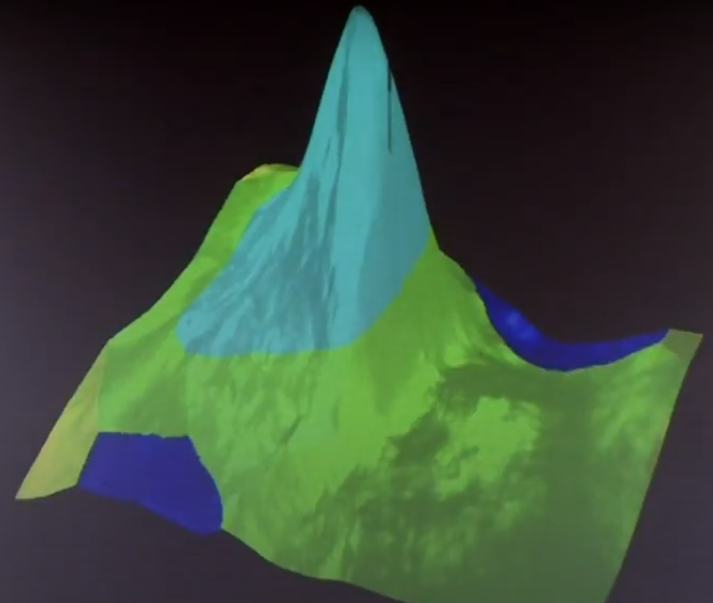}}
    \subfigure[Fig. 8b This is what the artificially made complex loss function of neural network looks like]{
    \includegraphics[width=0.5\linewidth,height=6cm]{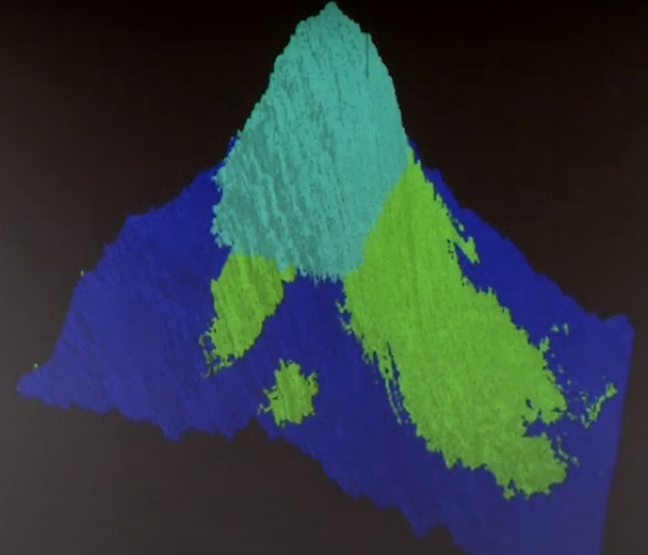}}
    \subfigure[Fig. 8c Zoomed version of 2nd figure]{
    \includegraphics[width=0.5\linewidth,height=6cm]{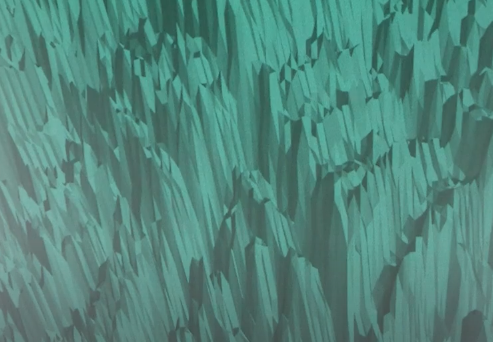}}
\end{figure}

The researchers made a defence technique whose loss function looks like this(fig. 8b) where the attacker cannot find in which direction to go for a gradient attack. If he starts at random point and proceeds in random direction, he won’t be able to figure out which direction to go and basically make no progress. Defence proposers on purpose came up with a loss function that is more complex so that attackers would not be to perform gradient descent as easily as they were able to do it before.

So the flaw in this defence model is that it only makes sure that it would take more time to for an attack to successfully take place but sooner or later the attacker might be able to attack successfully. Defence model should not have an aim to just make the task of adversary more time consuming. This is the reason why many of the defence model are broken in short time span.

\subsection{More iterations should be performed to check}

As presented in one of the paper(fig. 9) as they increase the number of iteration of gradient descent the accuracy of defence model goes on decreasing. If they had stopped at 1000 iterations they would be misled by the false results as it is approximately 7 times higher than what it is in one million iteration. So as the number of iterations increase the attack model becomes stronger. Sometimes it is necessary to go till millions of iterations.

  \begin{figure}
     \centering
     \includegraphics[width=100mm,scale=0.9]{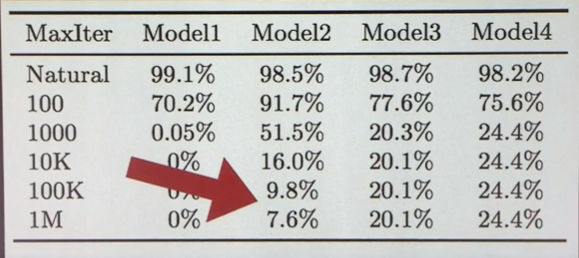}
     \caption{Why performing more iterations are necessary}
     \label{fig:my_label}
 \end{figure}

\subsection{Never underestimate hyperparameters}
Choice made while selecting the hyperparameters can have a large influence on the resulting model. Choices like activation function, pool size, kernel size, dropout, $\varepsilon$, and combination of these parameters play a very important role. The following paper [10] found out that not all parameters have discernible impact. For example, number of hidden layers and unit is found to have very small impact on the model. 

While comparing the models, the models should be trained on same optimizers. Without involving comparable hyperparameters in preparing, the models being looked at may have changed degrees of protection from adversarial models, so comparing the outcomes assembled from such defence models against various models might not be a wise choice.

All hyperparameters used must be clearly mentioned so that any future work if wants to build on foundation of current work would find it easy to do it.

\subsection{Try to break what you made}
If you are building a defence model for six months and have succeeded in defending the existing attacks, you must try out your own attack that makes your defence fail. Because believing that the attacker would not know what the defence looks like and how it works would not be appreciated, as this is a real case scenario as attacker would be knowing full defence model of your work as you have published a paper on it. 

\subsection{Actual Results should be claimed}

When claiming that your defence gives 95\% accuracy against the attack model that gives 92\% accuracy. But the attack model’s accuracy should be the recent one and not when it was published. As time progresses new models come and they keep on decreasing the accuracy of older models. So mentioning the latest accuracy would be appreciated. Moreover, defence model must be compared for same threat model. Comparing models evaluated on different threat model and then claiming that “A model is better than B” would not make much sense. 
Many attack have been shown on this site (https://www.robust-ml.org/) where the claims are significantly proven wrong. And latest figures and accuracy is provided.



\section*{Conclusion}
Learning from the past defence researches and the and finding the reason why some of them were broke easily and very sooner than expected, we have tried to provide some recommendation in form of checklist while building and evaluating the defence model. It mainly focuses on defence checklist against adversaries in image. These are not hard and fast rules to which are mandatory but if implemented some of them in positive way will increase the credibility of model and would prove that the defence model is robust enough.

Moreover, coming up with strong defence model would be of great boon to the machine learning community and folks directly or indirectly using ML in their day-to-day life. 

\section*{Future Works}
Researchers need to dive deep into interpretability of neural networks about why these robust and non-robust features are there, how we can circumvent them while making our neural network train. Why these models are overdependent on local features rather than global features of image. We want to make sure such simple adversaries don’t become blind spot for our machine learning models. We do not want neural networks to obtain the result and we consider it to be like black magic, unaware about what is going under the hood. Interpretable machine leaning is somewhat esoteric topic and efforts must be made dive deep into it. And most of the research in this area is been done on classification model. But making defences for GANs would also be appreciated. We know many product based companies to some extent are dependent on deep generative models. Suppose a big company is working on prediction of some life threatening disease, a data scientist extracts all the data from company’s database and trains a model which works pretty well, now to use this model for thousands of user’s products and to increase the performance we would need more data. So one would produce synthetic data using deep generative model, this is the best place where attacker would be all ready and had poisoned the data and this could enter your whole pipeline and the damage is done. The integrity and respectability of one’s company might be in danger.

\section*{References}
[1] N. Carlini, P. Mishra, T. Vaidya, Y. Zhang, M. Sherr, C. Shields,
D. Wagner, and W. Zhou, “Hidden voice commands,” in 25th USENIX
security symposium (USENIX security 16), 2016, pp. 513–530.

[2] J. Steinhardt, P. W. W. Koh, and P. S. Liang, “Certified defenses for data
poisoning attacks,” Advances in neural information processing systems,
vol. 30, 2017.

[3] J. Vincent, “Twitter taught microsoft’s ai chatbot to be a racist asshole in
less than a day,” The Verge, vol. 24, 2016

[4] A. Ilyas, S. Santurkar, D. Tsipras, L. Engstrom, B. Tran, and A. Madry,
“Adversarial examples are not bugs, they are features,” Advances in neural
information processing systems, vol. 32, 2019.

[5] K.-A. A. Wei, “Understanding non-robust features in image classifica-
tion,” Ph.D. dissertation, Massachusetts Institute of Technology, 2020.

[6] T. Liu, X. Xie, and Y. Zhang, “Zkcnn: Zero knowledge proofs for
convolutional neural network predictions and accuracy,” in Proceedings
of the 2021 ACM SIGSAC Conference on Computer and Communications
Security, 2021, pp. 2968–2985

[7] R. K. Bellamy, K. Dey, M. Hind, S. C. Hoffman, S. Houde, K. Kannan,
P. Lohia, J. Martino, S. Mehta, A. Mojsilovic et al., “Ai fairness 360: An
extensible toolkit for detecting, understanding, and mitigating unwanted
algorithmic bias,” arXiv preprint arXiv:1810.01943, 2018.

[8] Y. Dong, Q.-A. Fu, X. Yang, T. Pang, H. Su, Z. Xiao, and J. Zhu,
“Benchmarking adversarial robustness on image classification,” in Pro-
ceedings of the IEEE/CVF Conference on Computer Vision and Pattern
Recognition, 2020, pp. 321–331.

[9] L. Engstrom, A. Ilyas, and A. Athalye, “Evaluating and under-
standing the robustness of adversarial logit pairing,” arXiv preprint
arXiv:1807.10272, 2018.

[10] C. Burkard and B. Lagesse, “Can intelligent hyperparameter se-
lection improve resistance to adversarial examples?” arXiv preprint
arXiv:1902.05586, 2019.

\end{document}